\documentclass[12pt]{article}
\usepackage{graphicx}
\usepackage{bm}
\usepackage{amssymb}
\begin{document}
\begin{titlepage}

\vspace{1cm}

\begin{center}{\Large {\bf Generation of circularly polarized photons
\\for  a linear collider polarized positron source }}\\
\vspace{1.0cm} {\Large V.Strakhovenko$^a$} \footnote {Corresponding author, e-mail:
v.m.strakhovenko@inp.nsk.su }, {\Large X.Artru$^b$, R.Chehab$^{b,c}$, and
M.Chevallier$^b$,}\\
\vspace{1cm} {\it $^a$  Budker-INP, 11 Ac.Lavrentyeva, 630090, Novosibirsk, Russia\\
\vspace{0.5cm}$^b$ IPN-Lyon, IN2P3/CNRS et Univ. Claude Bernard, 69622 Villeurbanne,
France\\ \vspace{0.5cm} $^c$ LAL, IN2P3/CNRS et Univ. de Paris-Sud, BP 34-91898, Orsay
cedex, France}
\end{center}
\vspace{4.0cm}

\begin{abstract}

\noindent Various methods of obtaining longitudinally polarized positrons for future
linear colliders are reviewed . Special attention is paid to the schemes using circularly
polarized high-energy photons for positron production. Most effectively such photons are
obtained from electrons passing through a helical undulator or colliding with a
circularly polarized laser wave. Spectrum and polarization of radiation emitted during
helical motion of electrons are considered in detail. A new simple presentation of known
formulas is used to account for the influence of the wave intensity, of the electron-beam
angular divergence, of the collimation of radiation, and of the lateral and temporal
profiles of the laser bunch on the radiation properties.

\end{abstract}
\vspace{2cm}
\vspace{1cm}

\end{titlepage}
\newpage
\section{Introduction}
Systematic studies (see, e.g.\cite{Morg}) have shown that the use of polarized beams in a
$e^+ e^-$ linear collider will greatly help in the identification of new particles, in
search for new physics and in precision measurements of the coupling parameters. The
polarization of the electron beam alone is already very useful in this respect. If the
positron beam is also polarized, one can benefit from i) an increase of the effective
polarization: for instance, a 80$\%\,e^-$ polarization and a 60$\%\,e^+$ polarization
combine into an effective polarization of 94$\%$; ii) a better precision in the effective
and individual polarizations; iii) a further reduction of background events: for
instance, the number of $W^-W^+$ pairs is reduced by a factor 2 as compared to the case
when only the $e^-$ beam is polarized; iiii) an increased sensitivity to non-standard
couplings. In addition to these improvements, the transverse polarization of both $e^-$
and $e^+$ would allow one to investigate CP violating couplings.

The polarized electron sources have been pioneered at SLAC, since the mid-1970s
\cite{Ralley}, and used systematically for physics experiment since 1992. The SLC
electron beam was polarized to around 80$\%$. Such sources are based on a strained
semiconductor photocathode, which absorbs circularly polarized laser photons of energy
close to the band gap. State-of-the-art of polarized electron sources shows that they are
presently meeting the linear collider requirements.

All the schemes for obtaining longitudinally polarized positrons (except that using
$\beta^+$ decay of some isotopes) are based on the reaction $\gamma + \gamma \rightarrow
e^+ +e^-$, where at least one of two photons should be circularly polarized. When the
$\gamma$-conversion occurs in an amorphous target, the role of the second photon in the
reaction is played by an unpolarized Coulomb photon providing the momentum exchange
between the created particles and an atom. So, the incident photon should be circularly
polarized and have the energy, $\omega$, above the threshold value of 2$m$  ($m$ is the
electron mass, a system of units $\hbar=c=1$ is used ). The helicity transfer at high
energy was established in \cite{OlsMax} for two basic QED-processes: bremsstrahlung from
electrons and $e^+e^-$ pair production by photons. For both processes, the helicity
transfer is the most effective in the hard part of the spectra. The pair production cross
section $\sigma(\omega)$ increases with $\omega$, e.g., $\sigma(100 MeV)/\sigma(10
MeV)\simeq 2.44$ for tungsten. From this point of view, the higher is $\omega$ the larger
is the positron yield. However, due to a large phase space of produced positrons, they
can be successfully accepted by existing matching systems only if their energy does not
exceed several tens of MeV. Another argument, which may be used for a proper choice of
$\omega$, is to keep off the main nuclear resonances, thereby diminishing a harmful
hadronic background.

If polarized electrons are available, an amorphous target can be used according to
results of \cite{OlsMax} for subsequent emission of circularly polarized photons and
their conversion into polarized $e^+e^-$ pairs. Such a possibility was estimated in
\cite{Potyl} for electrons of energy $\varepsilon=50 MeV$ traversing a thin ($\sim 0.2
X_0$) amorphous target. Selecting positrons with $\varepsilon > 25 MeV$, a yield of
$2\cdot 10^{-3}$ per one initial electron was found. The mean polarization was about 0.6
of that of the initial electron beam.

When $e^+ e^-$ pairs are produced in collision of two real photons, the threshold
condition (for head-on collision) reads $\omega_h\cdot\omega_s\geqslant m^2$. Evidently,
within this scheme one of the photons is a circularly polarized laser photon, which,
typically, is rather soft (e.g., $\omega_s=0.117 eV$ for $CO_2$ laser and $\omega_s=2.23
eV$ for Nd:glass laser). Unpolarized hard photons may be obtained from high-energy
electrons radiating in amorphous or crystal targets. The latter option was considered in
\cite{Bcheh}. However, from the threshold condition, these photons should be really hard:
$\omega_h > 2.23 TeV $ for $CO_2$ laser and $\omega_h > 112 GeV $ for Nd:glass laser. In
turn, the electron energy should be appreciably larger than $\omega_h$.

The circularly polarized photons can be emitted by unpolarized electrons under helical
motion, which can be realized, in particular, in a helical undulator or in a circularly
polarized laser wave. The use of a helical undulator for the production of polarized
positrons was proposed first in \cite{BalMik}. Such technique requires incident electrons
of very high (of hundreds GeV) energy to produce a $\gamma$ beam of tens MeV, since the
undulator period is relatively large ($\sim$ 1 cm). This scheme is proposed in
\cite{Tesla} for TESLA LC. It is also considered for the NLC project and a dedicated
experiment \cite{Slac} is to be done at SLAC as a "proof of principle". One of the
challenges of this method, which requires a very long ($\sim$ 200 m) undulator, is an
accurate alignment.

A design of a polarized positron source for LC using  a circularly polarized laser wave
is proposed in \cite{Omori}. Here the challenges are concerning mainly the laser. Some
optimization for the total laser power is worked out and particular efforts are put on
the optics. Presently the needed number of $CO_2$ lasers is 10 and the incident beam
energy of 5.8 GeV is chosen, providing the maximum $\gamma$ energy of 60 MeV. One of the
advantages of this method is to have electron and positron main systems independent, as
the electron drive beam is of some GeV compared to the 250 GeV beam considered for TESLA.

Polarization of the emitted photons can be measured using Compton scattering of the
polarized photons in a magnetized iron \cite{Gold}. The same technique may be applied to
measure the polarization of created positrons after obtaining from them polarized
bremsstrahlung photons in an amorphous radiator.

In the present paper we consider the spectrum and polarization of photons emitted from
high-energy electrons under helical motion. The description of the phenomenon is
independent of the way how such motion is realized. This is due to the fact that the
physics of the process is completely determined by the type of trajectory. Particular
attention is paid to the practically important cases of a helical undulator and of a
laser wave, which are investigated in detail. We derive a new, rather simple but exact,
presentation of known formulas and use it to estimate the influence of the wave
intensity, of the electron-beam angular divergence, of the collimation of radiation, and
of the lateral and temporal profiles of a laser bunch on the radiation properties.

\section{Characteristics of radiation at helical motion of charged particles}
Let us, first, remind one some results of the so-called quasi-classical operator method
(QCOM, see e.g., \cite{book}), which are important here. Within this method, which
accurately takes into account recoil effects, the radiation characteristics are expressed
via the classical trajectory (velocity) of a charged particle in a given external field.
At appreciably large energies this is true for a wide class of external fields (see the
applicability conditions of QCOM in \cite{KatStr}). Thus, if different field
configurations provide similar electron trajectories, radiation will be similar as well,
being described by the same formulas. As was emphasized in \cite{BKS81}, one pair of such
equivalent systems is presented just by a helical undulator and a circularly polarized
laser wave. Really, in both cases the transverse momentum, $\bm {p}_{\perp}(t)$ is
\begin{equation}\label{ptrans}
\bm {p}_{\perp}(t)\,=\,<\bm {p}_{\perp}> + p_{\perp}\bigl[\bm {e}_1\cos(\omega_0t)+
\lambda\bm {e}_2\sin(\omega_0t)\bigr] \,\,,
\end{equation}
where $<\bm {p}_{\perp}>$ is a mean value of $\bm {p}_{\perp}(t)$, so that $<\ldots>$
means the averaging over a period of motion, and $\lambda=\pm 1$ corresponds to the sense
of rotation of $\bm {p}_{\perp}(t)$. We use the Cartesian basis ($\bm{e}_1,\bm{e}_2,
\bm{e}_3 $), where $\bm{e}_3 $ is directed along the electron beam momentum, i.e.,
parallel to the undulator axis. We suppose that a laser wave propagates towards the beam.
Then the "trajectory helicity", $\lambda$, in Eq.(\ref {ptrans}) coincides with that of
the magnetic field in an undulator and is opposite to the helicity of the laser wave.

Another argument in favor of the similarity under discussion was also given in
\cite{BKS81}. This is a similarity of the undulator electromagnetic field (UF) to the
field of a laser wave (LW) in the reference system moving along $\bm{e}_3 $ with
$V=<v_3>$, where an electron is on the average at rest (RS). Really, in the laboratory
system (LS), UF providing the trajectory (\ref {ptrans}) is purely magnetic and reads
\begin{equation}\label{undlab}
\bm {H}_{LS}\,=\,B\bigl[\bm {e}_1\cos\zeta+ \lambda\bm {e}_2\sin\zeta\bigr];\quad \zeta =
k z =(\varkappa x);\quad z=x^3  \,\,,
\end{equation}
where 4-vector $\varkappa=(0,0,0,-k)$, $k=2\pi/\lambda_u$, and $\lambda_u$ is the
undulator period. In the electron rest system, we obtain for UF
\begin{eqnarray}\label{undrest}
\bm {H}_{RS}^u &\!\!=&\!\! \frac{\varkappa'_0B}{k V}\bigl[\bm {e}_1\cos\zeta+ \lambda\bm
{e}_2\sin\zeta\bigr] \,\,,\\ \bm {E}_{RS}^u &\!\!=&\!\!\varkappa'_0\frac{B}{k}
\bigl[-\lambda\bm {e}_1\sin\zeta+ \bm {e}_2\cos\zeta\bigr]\,,\nonumber
\end{eqnarray}
where $\gamma=(1-V^2)^{-1/2}$ and $\varkappa'= \gamma k(V,0,0,-1)$, that is
$\varkappa'_0= \gamma kV \approx  \gamma k$. At $\gamma \gg 1$ the electromagnetic field
(\ref {undrest}) corresponds to the wave of quasi-photons ($0\neq \varkappa'^2 = k^2 \ll
m^2$) propagating towards the electron beam and having the helicity $-\lambda$. In turn,
LW with the helicity $-\lambda$ is described in LS by the vector potential
\begin{equation}\label{vecpot}
\bm {A}\,=\,a\bigl[\bm {e}_1\cos\chi+ \lambda\bm {e}_2\sin\chi\bigr],\quad \chi
=(\varkappa x),\quad \varkappa=\varkappa_0(1,0,0,-1) \,\,.
\end{equation}
Then in RS the LW-field reads
\begin{eqnarray}\label{waverest}
\bm {H}_{RS}^w &\!\!=&\!\!-\lambda a \varkappa'_0\bigl[\bm {e}_1\cos\chi+ \lambda\bm
{e}_2\sin\chi\bigr]
\,\,,\nonumber\\\\
\bm {E}_{RS}^w &\!\!=&\!\! -\lambda a \varkappa'_0\bigl[-\lambda\bm {e}_1\sin\chi+ \bm
{e}_2\cos\chi\bigr]\,.\nonumber
\end{eqnarray}
Here $ \varkappa'_0=\gamma\varkappa_0(1+V)\approx  2\gamma\varkappa_0$ is the laser
photon frequency (energy) in RS. The similarity of two fields in RS is clearly seen from
Eqs.(\ref {undrest}) and (\ref {waverest}).

Radiation in LW depends on two parameters. The intensity parameter, $\xi^2$, is expressed
via a vector potential: $\xi_w^2=e^2<\bm {A}^2>/m^2$ (e is the electron charge). An
expression for this parameter in the case of an undulator may be obtained from comparison
of equations (\ref {waverest}) and (\ref {undrest}) resulting in $\xi_u^2=(eB/km)^2$. (In
the literature dedicated to undulators, the notation $K^2$ is usually used for this
parameter.) Going back to Eq.(\ref {ptrans}), we conclude that the amplitude
$p_{\perp}=\xi m$, while $\omega_0=k$ for an undulator and $\omega_0=2 \varkappa_0$ for
LW. In fact, the argument of oscillating functions in Eq.(\ref {ptrans}) is $(\varkappa
x(t))$. In LS, the phase $(\varkappa x(t))$ goes over into $\omega_0t$ at assumed
conditions: when the electron energy is appreciably large,  $\varepsilon/m\gg 1$, and the
transverse velocity of the electron is small $v_{\perp}\ll 1$. The parameter $\xi^2$ can
be expressed also in terms of the particle trajectory $$\xi^2=\frac{<\bm
{p}_{\perp}^2(t)> - <\bm {p}_{\perp}(t)>^2 } {m^2}\,\,,$$ characterizing thereby the
deviation of the trajectory from a straight line. On the other hand, $\xi^2$ gives a
scale of the interaction strength (see, e.g., p.259 in  \cite{book}) in the photon
emission. In fact, we have roughly $\xi^2\approx\alpha N_{int}$, where $\alpha\simeq
1/137$ is the fine structure constant and $N_{int}$ denotes the number of LW photons (or
of quasi-photons in the undulator case) within the interaction volume, $V_{int}$ . The
latter is approximately $V_{int} \sim \lambda_C^2 \lambda_w$ since the transverse size of
the interaction region is about the Compton wave length, $ \lambda_C=1/m\simeq 3.86\cdot
10^{-11}cm$, and the longitudinal size is about the LW length, $\lambda_w$. As known, the
photon emission rate is roughly proportional to $\xi^2$ for $\xi^2 \lesssim 1$. From this
point of view, larger values of $\xi^2$ are more attractive. For a helical undulator,
where
$$\xi^2_u = [0.935\cdot B(T)\cdot \lambda_u(cm)]^2\,\,,$$ values of $\xi^2 \sim 0.5 \div
1$ seem achievable. In the laser case, extremely high power density $P$ is needed to
reach even $\xi^2 \sim 0.1$ as $$\xi^2_w = 3.66\cdot 10^{-19}\lambda_w^2(\mu m)\cdot
P(W/cm^2)\,\,.$$ For example, in \cite{Omori} a peak $P$ of $2.9\cdot10^{15}\,W/cm^2$ is
assumed at the focal point, which corresponds to the maximal value $\xi^2_{max}\simeq
0.12$. Note that $\xi^2_w$ diminishes at the periphery of a laser bunch proportional to
the density $n_w$ of laser photons ($\xi^2_w=2\alpha\lambda_w\lambda_C^2 n_w $).

Radiation in LW depends also on the purely kinematic parameter $s=2(\varkappa p)/m^2$,
which appears in the description of the conventional Compton scattering. Recollect that
the edge of the Compton spectrum is at $u\equiv\omega/(\varepsilon-\omega)=s$ (or at
$x\equiv\omega/\varepsilon=s/(1+s)$). We have $s_u\simeq 0.95\cdot 10^{-6}\cdot
\varepsilon(GeV)/\lambda_u (cm)$ for a undulator and $s_w\simeq 1.53\cdot 10^{-2}\cdot
\varepsilon(GeV)\varkappa_0(eV)$ for LW. We emphasize that $s_u\ll 1$ at any reasonable
electron energy $\varepsilon$. A magnitude of $s_w\sim 1$ can be easily achieved.
However, at $s_w\sim 1$ emitted photons will be too hard, since, as explained above,
photons in the energy range of several tens of MeV are needed for effective production of
positrons. To meet this condition, noticeably different electron energies should be used
in undulators and LW, though $s\ll 1$ in both cases. Since at small $s$ soft photons with
$\omega/\varepsilon\sim s \ll 1$ are mainly radiated, the recoil effect may be neglected
and the classical description of the process becomes valid.

\subsection{Spectral-angular distribution of radiation}
We can present the differential rate (probability per unite length or time) of photon
emission from a unpolarized electron in the following general form
\begin{equation}\label{probgen}
\frac{dN_{\gamma}}{d\Gamma dl} = \frac{1}{2}(A+\bm{B}\bm{\zeta})\equiv
\frac{A}{2}(1+\bm{\eta}\bm{\zeta}) \, ,\quad \bm{\eta}=\frac{\bm{B}}{A}\,,
\end{equation}
where $\bm{\eta} $ is the Stokes vector of emitted radiation and the auxiliary vector
$\bm{\zeta}$ ($|\bm{\zeta}|=1 $) describes the analyzing ability of some ideal photon
detector. When we consider the spectral-angular distribution, $d\Gamma = d\omega
d\bm{n}_{\perp}$, where $\bm{n}_{\perp}=\bm{k}_{\perp}/\omega$. The quantities $A$ and
$\bm{B}$ are obtained from the appropriate expression for the probability derived within
QCOM (e.g., from Eqs. (3) and (4) in \cite{Strakh}) if a dependence on time of the
transverse momentum (velocity) is specified. When such a dependence is given by Eq.(\ref
{ptrans}), we obtain, doing as in \cite{BKS81}
\begin{eqnarray}\label{ABcompl}
\Bigl ( A,\bm{B}\Bigr)\!\!&=&\!\!\frac{\alpha}{\pi}\sum\limits_{n=1}^{\infty}\Bigl (
a^{(n)},\bm{b}^{(n)}\Bigr)\delta(1+\xi^2+y^2-\frac{n}{\nu})\,\,;\nonumber \\ \nonumber\\
a^{(n)}\!\!&=&\!\!\frac{\xi^2\varphi (u)}{4}\Bigl [ J^2_{n-1}(Z)+J^2_{n+1}(Z)-
2J^2_{n}(Z) \Bigr ] -J^2_{n}(Z) \,;\nonumber \\\\b^{(n)}_2\!\!&=&\!\!
\frac{\lambda\xi^2\varphi (u)}{4}\cdot\frac{1+\xi^2-y^2}{1+\xi^2+y^2}\Bigl [
J^2_{n-1}(Z)-J^2_{n+1}(Z)\Bigr ]\,; \nonumber \\
\nonumber\\ b^{(n)}_1\!\!&=&\!\! - b^{(n)}_{lin}\sin2\tilde\phi\,,\qquad b^{(n)}_3= -
b^{(n)}_{lin}\cos2\tilde\phi\,,\nonumber \\
\nonumber\\ b^{(n)}_{lin}\!\!&=&\!\!J^2_{n}(Z)+\xi^2 \Bigl [J^2_{n}(Z)-
J_{n-1}(Z)J_{n+1}(Z)\Bigr ]\,.\nonumber
\end{eqnarray}
Here
\begin{eqnarray}\label{notation}
\nu &=& u/s\,\,,\qquad s=\frac{2(\varkappa p)}{m^2}\,\,,\qquad u=\frac{\omega}
{\varepsilon-\omega}\,\,,\qquad \varphi (u)= 1+u +\frac{1}{1+u}\,\,,\nonumber
\\\\ Z &=& 2\nu \xi y\,\,,\qquad y^2=(\gamma \bm{n}_{ef})^2\,\,,\qquad
\gamma=\varepsilon/m\,\,,\qquad \bm{n}_{ef}=\bm{n}_{\perp}-<\bm
{v}_{\perp}>\,\,,\nonumber
\end{eqnarray}
$J_{n}(Z)$ are the Bessel functions and the angle $\tilde\phi$ is the azimuth of the
vector $\bm{n}_{ef}$. Expression (\ref {ABcompl}) follows from Eqs. (3.5)-(3.9) in
\cite{BKS81} and coincides with Eq.(5.23) in \cite{book} if the latter is divided by
$\omega$ to pass on from the intensity in \cite{book} to the probability considered here.
The only generalization here is in retaining $<\bm {v}_{\perp}>=<\bm {p}_{\perp}>/
\varepsilon$, which was set to zero in \cite{BKS81} and  \cite{book}. The appearance of
the $\delta$-function in (\ref {ABcompl}) is due to the assumption that the number of
periods in a structure, $N_{per}=L/\lambda_{u,w}$ ($L$ is a length of the undulator or LW
bunch) is infinitely large. For large but finite $N_{per}$, we obtain (see e.g.,
Eq.(1.148) in \cite{book}) a set of narrow (width of $1/N_{per}$) peaks at the same
points $\nu(1+\xi^2+y^2)=n$ as in Eq.(\ref {ABcompl}).

When radiation is emitted along the mean electron velocity, we have $\bm{n}_{\perp} =
<\bm {v}_{\perp}>$ and $y=Z=0$. Then the sum over $n$ in Eq.(\ref {ABcompl}) reduces to
the only term with $n=1$. Moreover, only $J^2_{n-1}(Z)$ does not vanish and goes to
unity. As a result, the linear polarization vanishes for the forward direction, while the
circular polarization is maximal as $\eta_2\longrightarrow \lambda$. Then the dependence
of the probability on the parameter $\xi^2$ becomes linear except for the term $\xi^2$ in
the argument of the $\delta$-function. In this connection, let us point out that the
harmonic's number, $n$, is actually the difference between numbers of LW photons absorbed
from and re-emitted into the wave. From angular momentum conservation and some additional
consideration it can be shown that only $n=1$ is allowed in the forward direction for the
circularly polarized LW. To the same time, $\xi^2$ in the $\delta$-function reflects a
change of the mean longitudinal velocity due to the interaction with a wave as a
classical object, namely, $v_3^2(t)\simeq 1-\gamma^{-2}-\bm {v}_{\perp}^2(t)$ and
$<v_3^2>\simeq 1-(1+\xi^2)/\gamma^2$. That is why the characteristic angle of radiation,
$\vartheta_{ph}$, becomes $\vartheta_{ph}\sim \sqrt{1+\xi^2}/\gamma$. When $y$ increases,
the circular polarization diminishes. It vanishes at $y^2=1+\xi^2$ and changes its sign
at further increase of the emission angle (of the quantity $y$).

The first harmonic ($n=1$) dominates also for $\xi^2\ll 1$, where we obtain from Eq.(\ref
{ABcompl}) using the notation defined in Eqs.(\ref {ABcompl}) and (\ref {notation})
\begin{eqnarray}\label{CompAB}
\Bigl (A^{(C)},\bm{B}^{(C)}\Bigr)\!\!&=&\!\!\frac{\alpha\xi^2}{\pi}\Bigl (
a^{(C)},\bm{b}^{(C)} \Bigr)\delta(1+y^2-\frac{1}{\nu})\,\,;\qquad
b^{(C)}_{lin}=\nu(1-\nu)\,\,,\nonumber
\\\\a^{(C)}\!\!&=& \!\!\frac{\varphi (u)}{4}-\nu
(1-\nu)\,\,,\qquad b^{(C)}_2=\frac{\lambda\varphi (u)}{4}\cdot \frac{1-y^2}{1+y^2} =
\frac{\lambda\varphi (u)}{4}(2\nu-1)\,\,.\nonumber
\end{eqnarray}
As should be, Eq.(\ref{CompAB}) reproduces well known formulas obtained within the Born
approximation of the perturbation theory in $\xi^2$ for the Compton effect. The latter is
usually described in terms of a cross section which is obtained from
Eq.(\ref{CompAB})after dividing by the flux. More precisely, we have at $\xi^2\ll 1$
\begin{equation}\label{Interrel}
\frac{d\sigma^{(C)}}{d\Gamma} = \frac{4\pi\alpha\varepsilon}{m^2\xi^2(\varkappa p)}\cdot
\frac{dN_{\gamma}^{(C)}}{d\Gamma dl}\,\,.
\end{equation}

Using the following relation (see 8.442 in  \cite{Gradr}) $$J_n^2(z)=\sum
\limits_{k=n}^{\infty}(-1)^{k-n}\,\Bigl(\frac{z}{2}\Bigr)^{2k}\,
\frac{C_{2k}^{k-n}}{(k!)^2}\,\,;\qquad C_N^M = \frac{N!}{M!(N-M)!}\,,$$ we can expand the
quantities $ A,\bm{B}$ defined by  Eq.(\ref{ABcompl}) in powers of the parameter $\xi^2$
\begin{eqnarray}\label{ABseri}
\Bigl(A,\bm{B}\Bigr)\!\!&=&\!\!\frac{\alpha}{\pi}\xi^2\sum\limits_{n=1}^{\infty}
\delta(1+ \xi^2+y^2- \frac{n}{\nu})\sum\limits_{k=n}^{\infty}(-1)^{k-n}
\,\frac{C_{2k}^{k-n}(\nu\xi y)^{2(k-1)}}{[(k-1)!]^2}\Bigl
(a^{(k)},\bm{b}^{(k)}\Bigr)\,\,;\nonumber
\\\\
a^{(k)}\!\!&=&\!\!\frac{\varphi (u)}{4} -\Bigl(\frac{\nu y}{k}\Bigr)^2 \,,\qquad
b^{(k)}_2= \frac{\lambda\nu\varphi (u)}{4k} (1+\xi^2-y^2)
\,\,; \nonumber \\
\nonumber\\ b^{(k)}_1\!\!&=&\!\! - b^{(k)}_{lin}\sin2\tilde\phi\,,\qquad b^{(k)}_3= -
b^{(k)}_{lin}\cos2\tilde\phi\,\,,\qquad b^{(k)}_{lin}=\Bigl(\frac{\nu
y}{k}\Bigr)^2\Bigl[1+ \frac{\xi^2}{k+1} \Bigr ]\,.\nonumber
\end{eqnarray}
Such presentation of the probability (\ref{probgen}) essentially simplifies calculations
for $\xi^2\lesssim 1$, which is the most interesting region in applications. The formulas
obtained above should be averaged over $<\bm {v}_{\perp}>$ using a corresponding
distribution in the incident electron beam.

\subsection{Spectral characteristics of radiation}
In practice, radiation is always collimated. Then, to obtain the spectrum and
polarization of photons accepted by a collimator, we should take an integral of the
following type
\begin{equation}\label{Integral}
\int\limits_{\Omega_c}^{}d\bm{n}_{\perp}\,\int d\bm{\theta}_e
f(\bm{\theta}_e)\Bigl(A,\bm{B}\Bigr)\,,
\end{equation}
where $\Omega_c$ is the solid angle of a collimator, $f(\bm{\theta}_e)$ is the
distribution over $<\bm {v}_{\perp}>$ (we have introduced $\bm{\theta}_e\equiv <\bm
{v}_{\perp}>$) normalized by $\int d\bm{\theta}_e f(\bm{\theta}_e)=1$, and $ A,\bm{B}$
are defined by  Eq.(\ref{ABcompl}) or Eq.(\ref{ABseri}). Below we consider a round
collimator with the opening angle $\vartheta_{col}$ and axially symmetric distributions
$f(\bm{\theta}_e)\equiv f(\bm{\theta}_e^2)$. Remember that the quantities $ A,\bm{B}$
depend on the angles $\bm{n}_{\perp}$ and $\bm{\theta}_e$ only in the combination (see
(\ref{notation})) $\bm{n}_{ef}= \bm{n}_{\perp}-\bm{\theta}_e$. The azimuth of
$\bm{n}_{ef}$ enters only the linear polarization via $\sin2\tilde\phi$ and
$\cos2\tilde\phi$, while the value of $\bm{n}_{ef}^2$ is fixed by $\delta$-function (
$(\gamma \bm{n}_{ef})^2=y^2=n/\nu-(1+\xi^2)$). After the shift
$\bm{\theta}_e\longrightarrow \tilde{\bm{\theta}}_e+\bm{n}_{\perp}$ in
Eq.(\ref{Integral}) the integration over $\tilde{\bm{\theta}}_e$ becomes trivial under
our assumptions. Really, the linear polarization vanishes at integration over azimuth and
integration over $|\tilde{\bm{\theta}}_e|$ is carried out using $\delta$-function. As a
result, we have
\begin{eqnarray}\label{Spectr1}
\frac{d^2N_{\gamma}}{d\omega dl}=
\frac{A^{(\omega)}}{2}\Bigl(1+\bm{\eta}^{(\omega)}\bm{\zeta}\Bigr) \, ,\quad
\bm{\eta}^{(\omega)}=\frac{\bm{B}^{(\omega)}}{A^{(\omega)}} \, ;\quad
B^{(\omega)}_1=B^{(\omega)}_3=0\,;\nonumber \\\\
\Bigl(A^{(\omega)},B^{(\omega)}_2\Bigr)=\frac{\alpha\xi^2}{\gamma^2}\sum\limits_{n=1}^{\infty}
\Theta(X_n)\,F(X_n,\vartheta_{col})\sum\limits_{k=n}^{\infty}(-1)^{k-n}
\,\frac{C_{2k}^{k-n}(\nu\xi^2 X_n )^{(k-1)}}{[(k-1)!]^2}\Bigl
(a^{(\omega)},b^{(\omega)}_2\Bigr)\,\,;\nonumber
\\  \nonumber  \\
a^{(\omega)}=\frac{\varphi (u)}{4} -\frac{\nu X_n }{k^2} \,,\qquad b^{(\omega)}_2=
\frac{\lambda\varphi (u)}{4k} [2\nu(1+\xi^2)-n]\,,\qquad X_n= n-\nu(1+\xi^2)\,\,.
\nonumber
\end{eqnarray}
where $\Theta(X_n)$ is the step function: $\Theta(x)=1\,$ for $x>0$ and $\Theta(x)=0\,$
for $x<0\,$ , the variable $\nu \propto \omega $ is defined in (\ref{notation}). The
function $F(X_n,\vartheta_{col})$, which appears in integration over $d\bm{n}_{\perp}$ in
Eq.(\ref{Integral}), has the form
\begin{equation}\label{Fgeneral}
F(X_n,\vartheta_{col})=\int
d\bm{n}_{\perp}\Theta\Bigl(1-\frac{\bm{n}_{\perp}^2}{\vartheta_{col}^2}\Bigr)f\Bigl(
\bm{n}_{\perp} +\bm{e}\mu\vartheta_{col}\Bigr)\,\,,\qquad \mu=\frac{1}{\gamma
\vartheta_{col}}\sqrt{ \frac{X_n}{\nu}}\,\,,
\end{equation}
where $\bm{e}$ is an arbitrary unit vector in the transverse plane. When the angular
spread $\Delta_e$ in the electron beam is appreciably small, $\Delta_e \ll
\vartheta_{col}$, we can substitute $(\mu\vartheta_{col})^2$ for $\bm{n}_{\perp}^2$ in
the argument of $\Theta$-function in (\ref{Fgeneral}). Then, using the normalization
condition for the function $f$, we find out that  $F(X_n,\vartheta_{col})\longrightarrow
\Theta (1-\mu^2)$ at $\Delta_e \ll \vartheta_{col}$. In this case, the contribution of
the $n$~-~th harmonic to the spectrum is non-zero in the segment
\begin{equation}\label{interval}
\frac{1}{1+\xi^2+(\gamma \vartheta_{col})^2}\leqslant \frac{\nu}{n}\leqslant
\frac{1}{1+\xi^2}\,\,,\qquad \frac{\omega}{\varepsilon}=\frac{s\nu}{1+s\nu}\,\,.
\end{equation}

In the absence of collimation, which corresponds to the limit
$\vartheta_{col}\longrightarrow \infty$ in above formulas, we obtain the evident result,
$F(X_n, \vartheta_{col})=1$, being independent of $\Delta_e$. Generally speaking, in this
case it is more convenient to calculate the spectrum and the rate, $dN_{\gamma}/dl= \int
\limits_ {0}^{\varepsilon} d\omega A^{(\omega)}$, using simple integral representation
for these quantities (see e.g., Eq.(5.34) in \cite{book}). In particular, we have for
small $s\ll 1$
\begin{equation}\label{Ratesm}
\frac{dN_{\gamma}}{dl}\Bigl(s\ll 1\Bigr)\simeq \frac{\alpha (\varkappa p)\,
\xi^2}{\pi\varepsilon}\int \limits_{0}^{\infty}\,\frac{d\tau}{\tau^2}\cdot
\frac{\sin^2\tau-\tau^2\cos2\tau}{\tau^2+ \xi^2(\tau^2-\sin^2\tau)}\,\,.
\end{equation}
We emphasize that, for $s\ll 1$, the rate turns out to be independent of the electron
energy $\varepsilon$ as $(\varkappa p)/\varepsilon = \omega_0$ (remember that $\omega_0=
2\varkappa_0$ for LW). The first term in the expansion of the rate (\ref{Ratesm}) in
$\xi^2$ at $\xi^2 \ll 1$ is simply $W_0\equiv 2 \alpha  \xi^2(\varkappa p)/3\varepsilon $
and,
\begin{figure}[h]
\centering
\includegraphics[width=0.6\textwidth
]{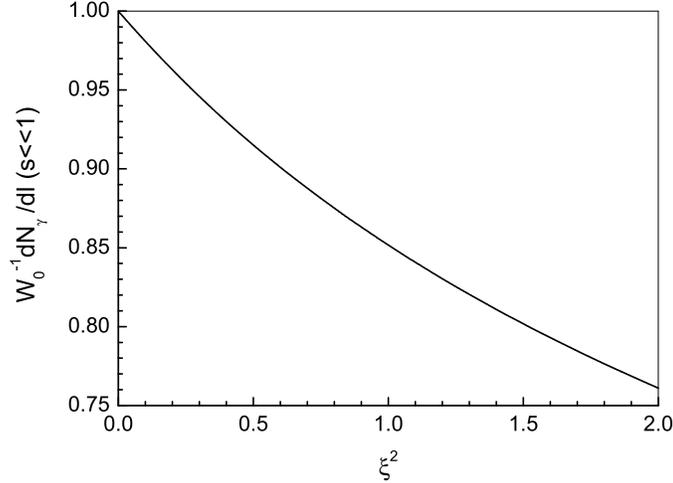}
\caption{Photon emission rate (\ref{Ratesm}) in units of  $W_0 $ . } \label{Fig:probksi1}
\end{figure}
using Eq.(\ref{Interrel}), the Thomson limit of the Compton-scattering cross section
$\sigma^{(C)}(s\ll 1)\simeq 8 \pi \alpha^2/3 m^2\simeq 665 mb $ is reproduced. The rate
(\ref{Ratesm}) is shown in Fig.\ref{Fig:probksi1} in units of  $W_0$

The intensity of radiation, $I_{\gamma}$, reads especially simple for $s\ll 1$ when the
classical formula is valid $$I_{cl}=I_{\gamma}(s\ll 1)= \frac{\alpha \xi^2 m^2 s^2}{6}
\equiv \frac{2}{3}\gamma^4 <w^2(t)>\,\,, $$ where $w(t)$ is the particle acceleration.
The radiative energy loss of an electron is determined by $I_{\gamma}$.

\begin{figure}[h]
\centering
\includegraphics[width=0.6\textwidth
]{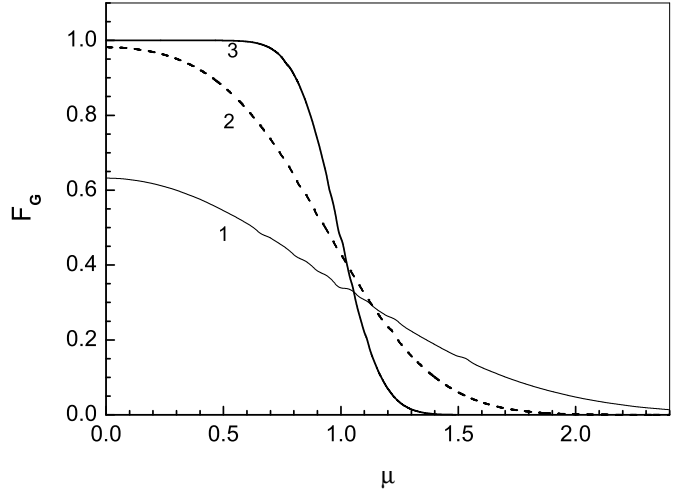}
\caption{Function  $F_G(\beta,\mu)$ at $\beta=1$ (1), $\beta=4$ (2), and $\beta=25$ (3) .
} \label{Fig:fbetmu2}\end{figure}

In further analysis of the role of photon collimation and angular divergence in the
electron beam we will use the Gaussian shape for the function $f$ $$f_G= \frac {1}{2\pi
\Delta_e^2} \exp\Bigl(- \frac{\bm{\theta}_e^2}{2\Delta_e^2} \Bigr)\,\,.$$ Then we obtain
from Eq.(\ref{Fgeneral})
\begin{eqnarray}\label{FGauss}
F_G(\beta,\mu)\!\!&=&\!\!2\beta \int\limits_{0}^{1} dx x\, \exp
[-\beta(x^2+\mu^2)]\,I_0(2\beta\mu x )=\nonumber
\\\!\!&=& \!\!\Theta(1-\mu^2)-\frac{1}{\pi}\int\limits_{0}^{\pi}\frac{d\phi}{g(\phi)}
(1+\mu \cos \phi) \exp [-\beta g(\phi)] \,\,;
\\\nonumber \\g(\phi)\!\!&=&\!\!1+2\mu \cos \phi+ \mu^2\,\,,\qquad  \mu^2= \frac{X_n}
{\nu (\gamma \vartheta_{col})^2}\,\,,\qquad \beta=\frac{\vartheta_{col}^2}{ 2\Delta_e^2}
\,\, ,\nonumber
\end{eqnarray}
where $I_0$ is the modified Bessel function. The first form of $F_G(\beta,\mu)$ in
Eq.(\ref{FGauss}) follows directly from Eq.(\ref{Fgeneral}), the second form makes more
obvious the general properties of $F(X_n,\vartheta_{col})$ discussed above. Shown in Fig.
\ref{Fig:fbetmu2} is $F_G(\beta,\mu)$ as a function of $\mu$ at different $\beta$.
As seen in Fig.\ref{Fig:fbetmu2}, the function
$F_G(\beta,\mu)$ is very like to that for a non-divergent beam already at $\beta=25$. At
$\beta=4$ the shape of the corresponding curve is substantially different from $\Theta
(1-\mu^2)$, but its value at $\mu=0$ (peak position) is still very close to 1. Finally,
at $\beta=1$ we have $F_G(\beta =1,\mu =0)\simeq 0.63$ and a slow decrease of $F_G$ when
$\mu$ increases. In other words, collimation of radiation within angles comparable with
the angular divergence of an electron beam is meaningless.

\begin{figure}[h]
\centering
\includegraphics[width=0.48\textwidth
]{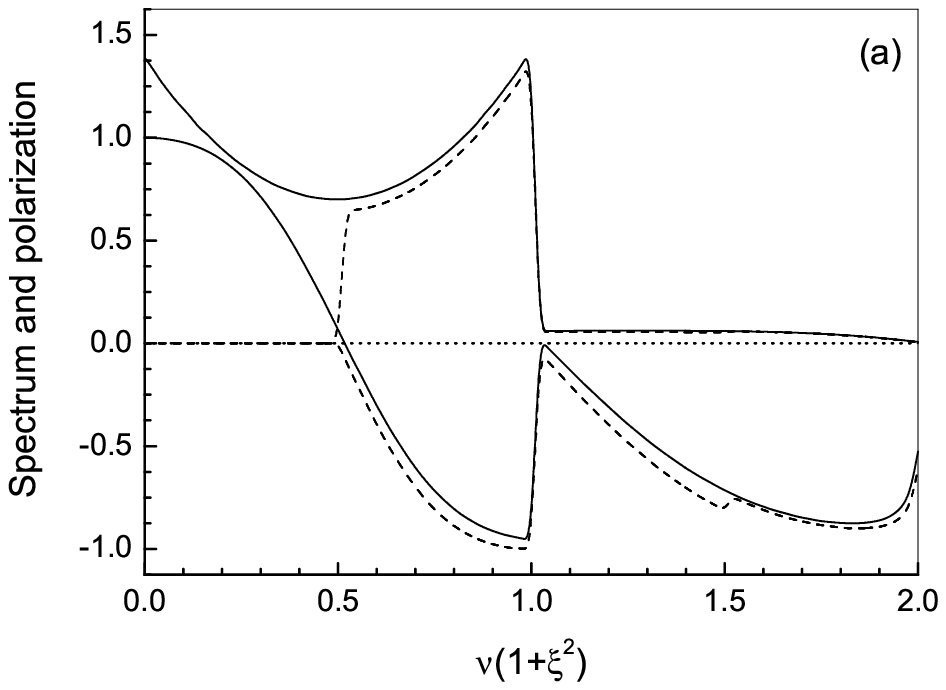} \hspace{0.025\textwidth}
\includegraphics[width=0.48\textwidth
]{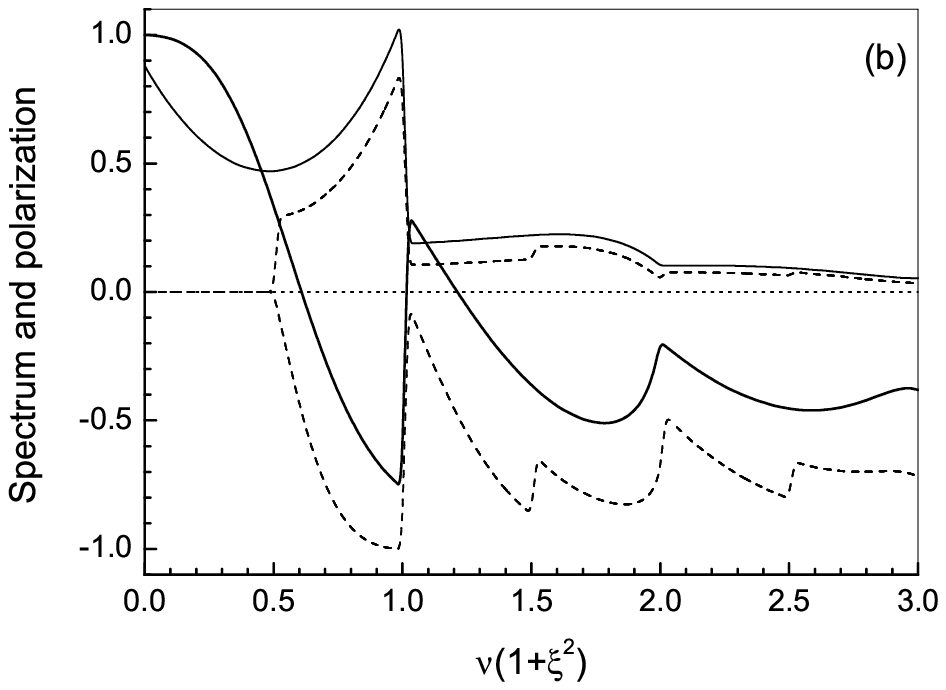} \caption{ For $\xi^2=0.1$ (a) and  $\xi^2=1$ (b); spectrum $W_{noc}^{-1}\cdot
d^2 N_{\gamma}/d\tilde {\nu}dl$ and polarization  $\eta_2^{(\omega)}$ in the absence of
collimation (solid), and at collimation within $\vartheta_{col}$ satisfying $(\gamma
\vartheta_{col})^2=1+\xi^2$ for a non-divergent electron beam (dashed). }
\label{Fig:ncoldiv3}
\end{figure}

\begin{figure}[h]
\centering
\includegraphics[width=0.48\textwidth
]{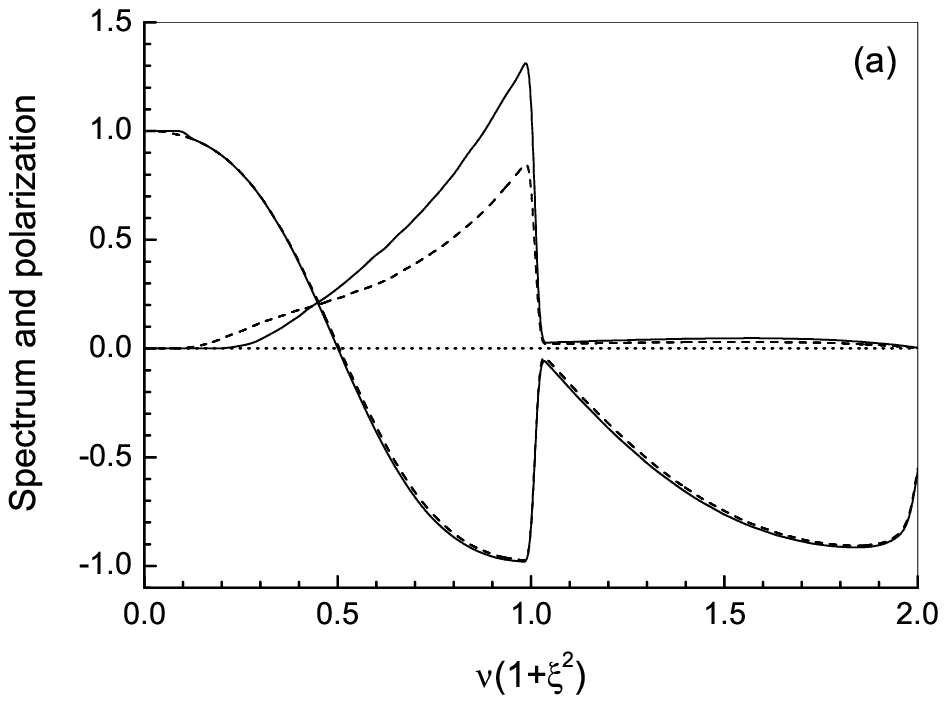} \hspace{0.025\textwidth}
\includegraphics[width=0.48\textwidth
]{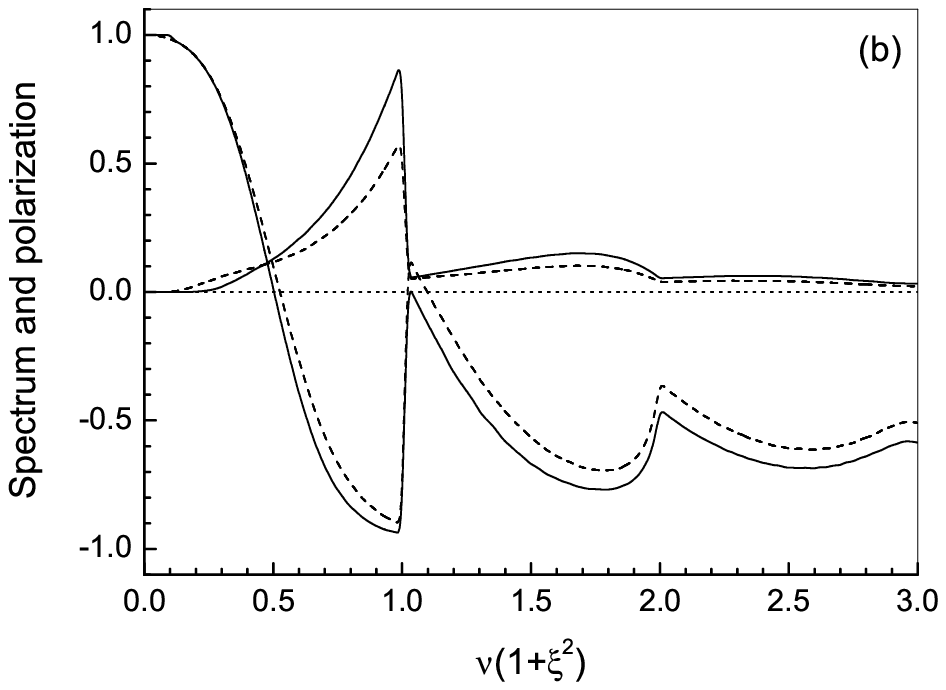} \caption{Same as in Fig.\ref{Fig:ncoldiv3} at the collimation within
$\vartheta_{col}$ satisfying $(\gamma \vartheta_{col})^2=1+\xi^2$ for a divergent
electron beam at $\beta=1$ (dashed) and $\beta=4$  (solid). } \label{Fig:coldiv4}
\end{figure}

The spectrum summed up over photon polarizations given by $A^{(\omega)}$ in
Eq.(\ref{Spectr1}), is plotted in Figs.\ref{Fig:ncoldiv3} and \ref{Fig:coldiv4} as a
function of $\tilde {\nu}=\nu(1+\xi^2)$. The component $\eta_2^{(\omega)}=
B^{(\omega)}_2/ A^{(\omega)}$ of the Stokes vector, which describes the circular
polarization of photons is presented as well. The kinematic parameter $s$ (see
Eq.(\ref{notation})) was set to $s=0.01$, when the normalization factor $W_{noc}$ is
given by Eq.(\ref{Ratesm}).Calculations are performed at $\lambda=-1$ (see
Eq.(\ref{ptrans}) for the definition of $\lambda$), which corresponds to the positive
helicity of LW and to the negative "helicity" of the electron trajectory in an undulator
or LW. Radiation spectra are presented for $\xi^2=0.1$ (a) and for $\xi^2=1$ (b). In
Fig.\ref{Fig:ncoldiv3}, radiation characteristics  in the absence of collimation (solid
curves) are compared with those at collimation within $\vartheta_{col}$ satisfying the
condition $(\gamma \vartheta_{col})^2=1+\xi^2$ for a non-divergent electron beam (dashed
curves).
At chosen collimation angle, the contribution of the n-th harmonic  (for a non-divergent
electron beam) is non-zero for $ n/2 \leqslant\tilde {\nu}\leqslant n$. This leads to
corresponding structures in collimated spectra, which are clearly seen in
Fig.\ref{Fig:ncoldiv3}. Since higher harmonics are more important
at $\xi^2=1$, reduction of the first peak height and increase of polarization degree are
more prominent in this case. A role of the beam divergence is illustrated by
Fig.\ref{Fig:coldiv4}, where spectra and polarizations are plotted for the same
$\vartheta_{col}$ as in Fig.\ref{Fig:ncoldiv3}, but at finite values of the parameter
$\beta$ (see Eq.(\ref{FGauss})). In accordance with Fig.\ref{Fig:fbetmu2}, structures at
$\tilde {\nu}=n/2$ are smeared out and additional suppression of the first peak at
$\beta=1$ is seen even for $\xi^2=0.1$.
\subsection{Total yield}
A total radiation yield during one pass of an electron through the whole undulator or a
single laser bunch is obtained by integration of the instantaneous characteristics over
time (length $l$). The electron energy $\varepsilon$ diminishes due to the radiative
energy loss given by the radiation intensity $I_{\gamma}$. In Eq.(\ref{Spectr1}),
$\varepsilon$ is represented explicitly ($\gamma^{-2}$ as a common multiplier and in the
quantity $\mu$ entering $F(X_n,\vartheta_{col})$) and implicitly via $\nu$ and $u$.
However, the total energy loss and the energy spread in the electron beam is typically
small as compared with $\varepsilon$ and therefore will be neglected below. In the
opposite case of the appreciably large energy loss, total radiation should be described
in terms of the electron-photon shower, which development is determined by the
probabilities obtained above. Strictly speaking, the collimation angle $\vartheta_{col}$
also depends on $l$. Practically this is important only in the case of an undulator when
the total length, $L$,  is comparable with the distance to a target. In the case of LW we
can neglect a variation of $\vartheta_{col}$ over $L$ as a laser bunch is usually very
short. The parameter $\xi^2$ is constant for an undulator and depends on $l$ for LW.

Thus, the total spectral yield from an undulator, $\,dN_{\gamma}/d\omega$, is given
within our approximation by Eq.(\ref{Spectr1}), if we multiply the quantities
$\,A^{(\omega)}, B^{(\omega)}_2$ by $L$ and substitute $\Phi = L^{-1}\,\int
\limits_{0}^{L}dl\,F(X_n, \vartheta_{col}(l)) $  for $F(X_n,\vartheta_{col})$. In fact,
we should perform the integration of $\Theta$-function in Eq.(\ref{Fgeneral}):
\begin{equation}\label{Totund}
\int \limits_{0}^{L}\frac{dl}{L}\Theta\Bigl(1-\frac{\bm{n}_{\perp}^2}{\vartheta_{col}^2
(l)}\Bigr) = \Theta\Bigl(1-\frac{\bm{n}_{\perp}^2}{\vartheta_{in}^2}\Bigr)+
\Theta\Bigl(1-\frac{\bm{n}_{\perp}^2}{\vartheta_{out}^2}\Bigr)\Theta\Bigl(\frac{\bm{n}_
{\perp}^2}{\vartheta_{in}^2}-1\Bigr)\Bigl[\frac{\vartheta_{out}}{|\bm{n}_{\perp}|}-1
\Bigr]\frac{\vartheta_{in}}{\vartheta_{out}-\vartheta_{in}} \,\,,
\end{equation}
where $\vartheta_{in}=\vartheta_{col}(0)$ and $\vartheta_{out}=\vartheta_{col}(L)$ are
the angles ($\vartheta_{out}>\vartheta_{in}$) at which a collimator is seen from the
entrance and the exit of the undulator, respectively. Using this result, the integration
over $\bm{n}_{\perp}$ in Eq.(\ref{Fgeneral}) can be performed. In particular, we obtain
for the Gaussian type of $f$ in Eq.(\ref{Fgeneral})
\begin{eqnarray}\label{TotGs}
\Phi_G\!\!&=&\!\!2Q\Bigl\{\int\limits_{0}^{\mu_{in}^{-1}} dx x\, \exp [-Q(x^2+1)]\,I_0(2Q
x )+\int\limits_{\mu_{in}^{-1}}^{\mu_{out}^{-1}} dx
\frac{1-x\mu_{out}}{\mu_{in}-\mu_{out}}\, \exp [-Q(x^2+1)]\,I_0(2Q x )\Bigr\}\nonumber
\\\Phi_{nod}\!\!&=& \!\!\Theta(1-\mu_{in}^2)+ \Theta(1-\mu_{out}^2)\Theta(\mu_{in}^2-1)
\frac{1-\mu_{out}}{\mu_{in}-\mu_{out}}\,\,;
\\\nonumber \\\mu_{in}\!\!&=&\!\!\frac{1}{\gamma
\vartheta_{in}}\sqrt{ \frac{X_n}{\nu}}\,\,,\qquad\mu_{out}=\frac{1}{\gamma
\vartheta_{out}}\sqrt{ \frac{X_n}{\nu}} \,\,,\qquad Q=\frac{X_n}{2 \nu
(\gamma\Delta_e)^2} \,\, .\nonumber
\end{eqnarray}
Here $\Phi_{nod}$ represents the limit of $\Phi_G$ at $Q\longrightarrow \infty$ which
corresponds to the non-divergent beam. Note also that the first item in $\Phi_G$
coincides with $F_G(\beta_{in},\mu_{in})$ from Eq.(\ref{FGauss}) and that
$Q=\beta_{in}\mu_{in}^2=\beta_{out}\mu_{out}^2 $ is independent of the collimation angle.
In the absence of the external collimator its role is played by the aperture of an
undulator. Thus, radiation which escapes the undulator may be described using $\Phi_G$
from Eq.(\ref{TotGs}), where we should set $\vartheta_{in}=\arctan(r/L)\sim r/L$ ($r$ is
the aperture radius) and  tend $\mu_{out}$ to zero.

As explained above, the electron energy and collimation angle may be considered as
constant during collision of an electron beam with a laser bunch. Thus, a variation of
the parameter $\xi^2$ only should be taken into account when obtaining the total yield.
Moreover, since $\xi^2$ depends also on coordinates $\bm{r}$ a convolution should be
performed with a corresponding distribution $n_e(\bm{r},t)$ in the electron beam. Let
$G(\xi^2)$ is some instantaneous characteristics of radiation, for example, $G(\xi^2)=
A^{(\omega)}$ from Eq.(\ref{Spectr1}), if we consider the spectrum. Then the
corresponding total yield, $G_{out}$, reads
\begin{equation}\label{Outlas}
G_{out}=\int\limits_{-\infty}^{\infty}dt\,\int d\bm{r}\, n_e(\bm{r},t
)\,G\Bigl(\xi^2(\bm{r},t)\Bigr) \,.
\end{equation}
This expression greatly simplifies for $\xi^2\ll 1$ when (see, e.g, Eq.(\ref{CompAB}))
the rather complicated, non-linear dependance of $G(\xi^2)$ on $\xi^2$ reduces to linear
one: $G(\xi^2\ll 1)\simeq \xi^2 G_0$, where $G_0$ is independent of $\xi^2$. As explained
above, we really have $\xi^2\ll 1$ for lasers. Then, recollecting the interrelation
between $\xi^2$ and the laser photon density $n_w$, which reads $\xi^2=4\pi \alpha n_w
/(\varkappa_0 m^2)$, we can rewrite Eq.(\ref{Outlas})
\begin{equation}\label{OutComp}
G_{out}^{(C)}=\frac{4\pi \alpha G_0}{\varkappa_0
m^2}\int\limits_{-\infty}^{\infty}dt\,\int d\bm{r}\, n_e(\bm{r},t )\,n_w(\bm{r},t ) \,.
\end{equation}
For Gaussian shape of the lateral profiles in both electron and laser bunches we have
\begin{eqnarray}\label{Latprof}
n_e(\bm{r},t )\!\!&=&\!\!N_e g_e\Bigl(z-z_c^{(e)}(t)\Bigr)\frac{\exp\Bigl(-
\bm{\rho}^2/2\sigma_e^2 \Bigr)}{2\pi \sigma_e^2} \,\,,\quad z_c^{(e)}(t)=z_0^{(e)}+t\,\,,
\quad z_c^{(w)}(t)=z_0^{(w)}-t\,\,, \nonumber
\\\nonumber\\ n_w(\bm{r},t ) \!\!&=&\!\!N_w g_w\Bigl (z-z_c^{(w)}(t)\Bigr)\frac{\exp\Bigl
(- \bm{\rho}^2/2\sigma_w^2(z)\Bigr)}{2\pi \sigma_w^2(z)} \,\,, \qquad
\int\limits_{-\infty}^{\infty}dz\,g_{e,w}(z)=1\,\,,
\end{eqnarray}
where $N_e$ and $N_w=E_b/\varkappa_0$ ($E_b$ is the laser-bunch energy) are the total
numbers of electrons and photons in corresponding bunches. Functions $g_{e,w}(z)$
characterize the longitudinal (temporal) profiles of the bunches and $z_c^{(e,w)}(t)$
mark positions of the bunch centers. Focusing of light, if any, is taken into account by
the $z$-dependent width $\sigma_w(z)$. The focal point corresponds to $z=0$. Evidently,
the yield is maximal when the bunch centers meet just at the focal point, i.e, when
$z_0^{(w)}+z_0^{(e)}=0$. Using densities (\ref{Latprof}) and making the shift
$t\longrightarrow t-z+z_0^{(w)}$, we obtain from Eq.(\ref{OutComp}) for the yield per one
collision
\begin{equation}\label{OutCGtr}
G_{out}^{(C)}=\frac{2\alpha N_e E_b G_0}{(\varkappa_0
m)^2}\int\limits_{-\infty}^{\infty}dt\,g_w(t)\,\int\limits_{-\infty}^{\infty}dz \frac{
g_e(2z-t)}{ \sigma_e^2+\sigma_w^2(z)} \,.
\end{equation}
Remember that for undulators we have presented the yield per one electron. Let $L$ be the
total length of the laser pulse. At focusing, the width $\sigma_w(z)$ increases rather
fast with growing $|z|$ , determining a $z$-interval, which contributes to the integral
in Eq.(\ref{OutCGtr}). When focusing is hard, the size of this interval, $L_{ef}$,
becomes noticeably smaller than $L$. Concerning the electron-bunch length, the shorter it
is, the higher the yield. Anyway, this length should be chosen less or about $L_{ef}$.
Diminishing of $\sigma_e$ in Eq.(\ref{OutCGtr}) increases the yield, the other parameters
being constant. However, we should bear in mind that $G_0$ depends on $\beta=
\vartheta_{col}^2/(2 \Delta_e^2) $ (see Eq.(\ref{FGauss})) and that the product $\Delta_e
\cdot \sigma_e $ has a lower bound given by the beam emittance. As shown above, $\beta$
should be appreciably large (small $\Delta_e$ are needed) to achieve the maximal value of
$G_0$ at given collimation conditions. Thus, both $\Delta_e$ and $\sigma_e$ should be
small enough, i.e, a low-emittance electron beam is needed.

\begin{figure}[h]
\centering
\includegraphics[width=0.6\textwidth
]{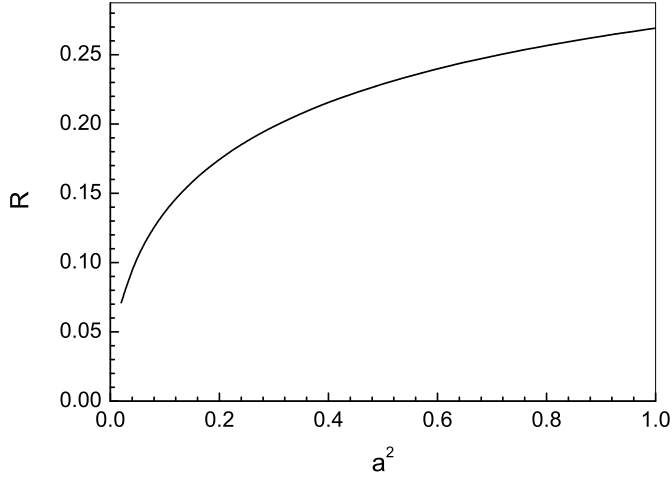}
\caption{Dependence of $R$ on $a^2$ according to Eq.(\ref{OutJap}) . }
\label{Fig:rasqur5}\end{figure}

To perform  further calculations in Eq.(\ref{OutCGtr}), we should specify a shape of the
functions $g_w(t)$ and $\sigma_w^2(z)$. Focusing will be taken into account by choosing
\begin{equation}\label{Sigmaz}
\sigma_w^2(z)=\sigma_w^2(0)[1+C_f y^2]\,,\,y=4z/L\,\,.
\end{equation}
Let us use for the illustration the temporal profile of a laser bunch considered in
\cite{Omori}, where
$$g_w(t)=\frac{4}{3L}\Bigl[\Theta(\tau+2)\Theta(-\tau-1)(\tau+2)+\Theta(\tau+1)\Theta(1-\tau
)+\Theta(2-\tau)\Theta(\tau-1)(2-\tau)\Bigr] \,\,,$$ with $\tau=4t/L$. Then, making the
shift $z\longrightarrow z+t/2$, we can take the integral over $t$ in Eq.(\ref{OutCGtr})
\begin{eqnarray}\label{OutJap}
G_{out}^{(C)}\!\!&=&\!\!\frac{8\alpha N_e E_b G_0 R}{3(\varkappa_0 m)^2 [\sigma_e^2
+\sigma_w ^2(0) ]}\,,\,\,\, R=a^2\int\limits_ {-\infty}^{\infty}dz
g_e(2z)\,\Psi(z)\,,\,\,\, \Psi(z)=\frac{1}{2}\ln
\Bigl[\frac{(1+d_3^2)(1+d_4^2)}{(1+d_1^2)(1+d_2^2)}\Bigr] +\nonumber
\\\\&+&d_1\arctan (d_1)+d_2\arctan (d_2)-d_3\arctan (d_3)-d_4\arctan (d_4)\,\,;\quad d_1=(1+y)/a
\,,\nonumber \\ \nonumber\\ d_2\!\!&=&\!\!(1-y)/a\,,\quad
d_3=\Bigl(\frac{1}{2}+y\Bigr)/a\,,\quad d_4= \Bigl(\frac{1}{2}-y\Bigr)/a\,;\quad
a^2=\frac{\sigma_e^2+\sigma_w^2(0)}{C_f \sigma_w^2(0)}\,,\quad y=4\frac{z}{L}\,.\nonumber
\end{eqnarray}
The quantity $R$, which was calculated using $g_e(z)=\exp(-z^2/2 \sigma_z^2)/\sqrt{2\pi
\sigma_z^2}$, is shown in Fig.\ref{Fig:rasqur5} as a function of $a^2$ at $r=L/\sigma_z=3
$ (this value of $r$ was used in \cite{Omori}). At given $a^2$ the quantity $R$ increases
with $r$.
 We found out that the lateral profile of a laser beam
in \cite{Omori}, which actually is doughnut-shaped (see Fig.10 in \cite{Omori}), can be
satisfactorily approximated in the region around the focal point by the Gaussian one with
$\sigma_w(0)=19\mu m$ and $C_f\simeq 11$ (see Eq.(\ref{Sigmaz})). Using also $\sigma_e=
20\mu m$, we obtain $a^2=0.192$ and, correspondingly, $R=0.172$. Now any spectral
characteristic of radiation may be estimated for the conditions of \cite{Omori} using
Eq.(\ref{OutJap}). For example, to estimate the total number of photons  emitted during a
single collision,  $G_0=2 \alpha (\varkappa p)/3\varepsilon$) should be substituted into
Eq.(\ref{OutJap}). In \cite{Omori}, the electron beam subsequently collides with laser
bunches at 200 collision points to produce a sufficient number of photons. The whole
length of the structure is about $60\,m$ and there are numerous mirrors and collimators
inside.
\begin{figure}[h]
\centering
\includegraphics[width=0.48\textwidth
]{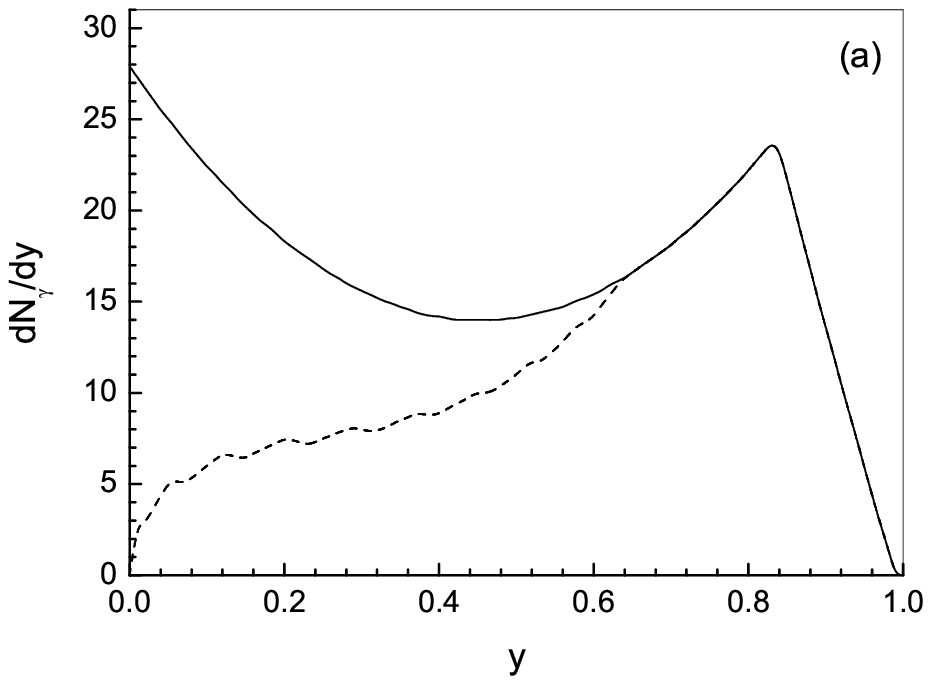} \hspace{0.025\textwidth}
\includegraphics[width=0.48\textwidth
]{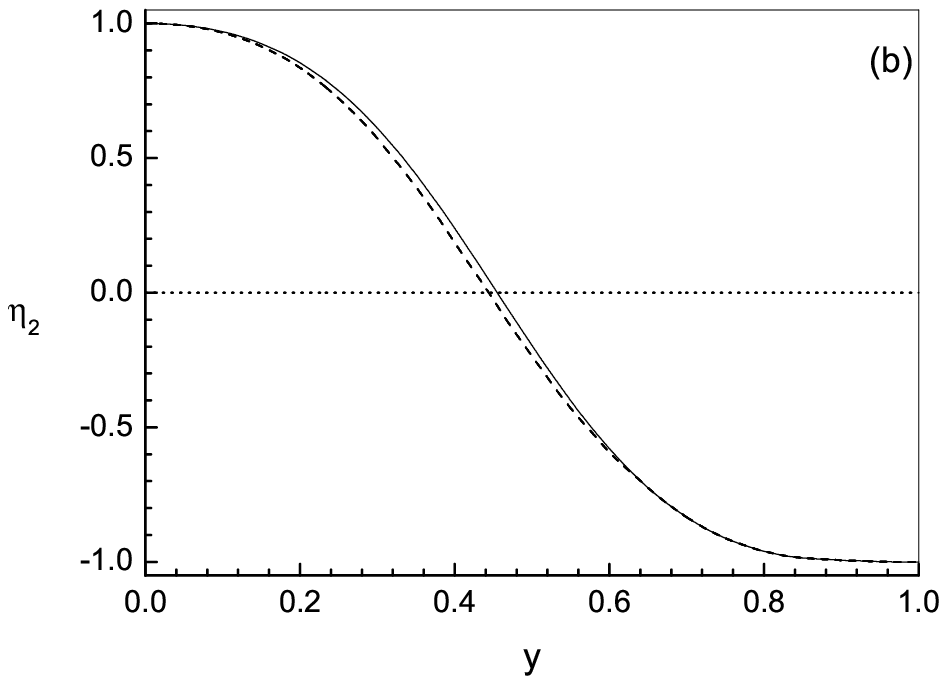} \caption{Spectrum  $d N_{\gamma}/dy$ and polarization $\eta_2^{(\omega)}$
for all generated photons (solid) and for photons which reach the target (dashed); $y=
\omega/\varepsilon_0/s_0$, $\varepsilon_0=5.8 GeV$, $s_0\equiv s(\varepsilon_0)\simeq
0.01$. } \label{Fig:japspec6}
\end{figure}
The yield from the whole structure on the $e^+e^-$-production target is a sum of the
individual yields, which are described by Eq.(\ref{Outlas})(or by its simplified version
(\ref{OutJap})). Note that the contribution from each collision point should be
calculated at its own collimation angle. The latter is determined by a distance between
the collision point and the last (the nearest to the target) collimator. Shown in
Fig.\ref{Fig:japspec6} are  spectra (normalized per one electron) and circular
polarization for all generated photons (solid) and for photons which reach the target
(dashed) for conditions of \cite{Omori}. The calculation is performed using
Eq.(\ref{OutJap})) and taking into account diminishing of the electron energy and the
change of the collimation angle, while a slow increase of $\sigma_e $ over a collision
section and the laser-power loss caused by mirrors are neglected. Comparison of our
results with Fig.14(b) in \cite{Omori} shows that even in its simplest form our approach
provides rather accurate estimations. For example, our calculation overestimate the total
number of all generated photons by $\sim 2.4\%$, and that for photons hitting the target
by $\sim 2.1\%$.
\section{Conclusion}
Among various methods of obtaining longitudinally polarized positrons for future linear
colliders, the most promising schemes are those using circularly polarized, high-energy
photons for positron production. Most effectively such photons are emitted from electrons
passing through a helical undulator or colliding with the circularly polarized laser
wave. Though the physics of the photon-emission process from electrons is the same in
both cases, a drastic difference in the periods of motion leads to the corresponding
difference in electron energy needed to produce positrons of several tens of MeV, which
can be effectively captured and accelerated. While an electron energy of several GeV is
appropriate in the laser case, hundreds of GeV are needed for undulators. Another
advantage of the laser scheme is in easy switching of the photon (and thereby positron)
helicity. This option is very useful in experiments. It seems, however, that there are
less questions in the construction of undulators, while producing of laser systems with
record parameters (very high power, high repetition rate) and rather sophisticated optics
is still a challenge. We propose a new, rather simple presentation of known formulas
describing the radiation properties in both cases. Our consideration takes into account
such factors as a magnitude of the wave intensity, angular divergence in the electron
beam, collimation of radiation, and the lateral and temporal profiles of a laser bunch.
The developed description allows one to choose easily a set of parameters optimizing the
photon yield.

 \vspace{0.25 cm} \noindent{\bf Acknowledgements} \vspace{0.25 cm}

\noindent One of us (V.S.) is thankful for kind hospitality during his stay at IPN~-~Lyon
where a part of this work has been done. He is also grateful to the Russian Fund of Basic
Research for partial support of this work by the Grant 03-02-16510.



\begin{thebibliography}{99}

\bibitem{Morg} G. Moortgat-Pick and H.Steiner, EPJdirect {\bf C6} (2001)1

\bibitem{Ralley}  R.Alley et al., {\em The Stanford linear accelerator polarized electron
source}, SLAC-PUB-95-6489, (1995).

\bibitem{OlsMax} H.Olsen and L.C.Maximon, Phys. Rev. {\bf 114} (1959) 887.

\bibitem{Potyl} A.P.Potylitsin, Nucl.Instrum.Methods Phys. Res.
A~{\bf 103} (1997) 395.

\bibitem{Bcheh} V.N.Baier, R.Chehab and V.M.Katkov, Nucl.Instrum.Methods
 Phys. Res. A~{\bf 338} (1994) 156.

\bibitem{BalMik} V.E.Balakin, A.A.Mikhailichenko, {\em The conversion system for obtaining
high polarized electrons and positrons}, INP 79-85, Novosibirsk, (1979).

\bibitem{Tesla} K. Fl$\ddot{o}$ttmann, {\em Investigations toward the development of
polarized and unpolarized high intensity positron sources for linear colliders}, DESY
93-161a, (1993).

\bibitem{Slac} A.W. Weidemann, {\em Polarized positrons at a future linear collider and
the final focus test beam }, SLAC-PUB-10581, (2004).

\bibitem{Omori} T.Omori, T.Aoki, K.Dobashi et al., Nucl.Instrum.Methods
 Phys. Res. A~{\bf 500} (2003) 232.

\bibitem{Gold} M.Goldhaber, L.Grodzins, A.Sunyar. Phys.Rev.{\bf 106} (1957) 826.

\bibitem{book} V.N. Baier, V.M. Katkov, and V.M. Strakhovenko,
{\em Electromagnetic Processes at High Energies in Oriented Single Crystals }, World
Scientific Publishing Co, Singapore, 1998.

\bibitem{KatStr}V.M. Katkov and V.M. Strakhovenko,
JETP {\bf 92} (2001) 561.

\bibitem{BKS81} V.N.Baier, V.M.Katkov and V.M.Strakhovenko,
JETP B~{\bf 53} (1981) 688.

\bibitem{Strakh} V.M. Strakhovenko, Phys. Rev. A~{\bf 68} (2003) 042901


\bibitem{Gradr} I.S. Gradshteyn and I.M. Ryzhik, {\em Table of Integrals, Series,
and Products}, 4th Edn. Academic Press, New York, (1965)



\end{thebibliography}
\end{document}